\documentclass[sigconf]{acmart}
\AtBeginDocument{%
  }

\usepackage{graphicx}
\usepackage{float}
\usepackage{tabularx}
\usepackage{makecell}
\usepackage{amsmath}

\usepackage{amssymb}
\usepackage{makecell}
\usepackage{booktabs}
\usepackage{multirow}
\usepackage{threeparttable}
\usepackage{todonotes}
\usepackage{placeins}

\usepackage{fancyhdr}
\renewcommand\footnotetextcopyrightpermission[1]{}
\acmConference{}{}{}{}
\begin{document}

%%
%% The "title" command has an optional parameter,
%% allowing the author to define a "short title" to be used in page headers.
\title{Retentive Relevance: Capturing Long-Term User Value in Recommendation Systems}

\author{Saeideh Bakhshi}
\email{bakhshi@meta.com}
\affiliation{%
  \institution{Meta}
  \city{Menlo Park}
 \state{California}
  \country{USA}
}

\author{Phuong Mai Nguyen}
\email{pmnguyen@meta.com}
\affiliation{%
  \institution{Meta}
  \city{Menlo Park}
 \state{California}
  \country{USA}
}

\author{Robert Schiller}
\email{rschiller@meta.com}
\affiliation{%
  \institution{Meta}
  \city{New York}
  \state{New York}
  \country{USA}
}
\author{Tiantian Xu}
\email{tiantianx@meta.com}
\affiliation{%
  \institution{Meta}
  \city{New York}
  \state{New York}
  \country{USA}
}

\author{Pawan Kodandapani}
\email{pawansk@meta.com}
\affiliation{%
  \institution{Meta}
  \city{Boston}
  \state{Massachusetts}
  \country{USA}
}

\author{Andrew Levine}
\email{andrewlevine@meta.com}
\affiliation{%
  \institution{Meta}
  \city{San Francisco}
 \state{California}
  \country{USA}
}

\author{Cayman Simpson}
\email{cayman@meta.com}
\affiliation{%
  \institution{Meta}
  \city{Menlo Park}
 \state{California}
  \country{USA}
}

\author{Qifan Wang}
\email{wqfcr@meta.com}
\affiliation{%
  \institution{Meta}
  \city{Menlo Park}
 \state{California}
  \country{USA}
}

%%
%% The abstract is a short summary of the work to be presented in the
%% article.
\begin{abstract}
  Recommendation systems have traditionally relied on short-term engagement signals, such as clicks and likes, to personalize content. However, these signals are often noisy, sparse, and insufficient for capturing long-term user satisfaction and retention. We introduce Retentive Relevance, a novel content-level survey-based feedback measure that directly assesses users’ intent to return to the platform for similar content. Unlike other survey measures that focus on immediate satisfaction, Retentive Relevance targets forward-looking behavioral intentions, capturing longer term user intentions and providing a stronger predictor of retention. We validate Retentive Relevance using psychometric methods, establishing its convergent, discriminant, and behavioral validity. Through large-scale offline modeling, we show that Retentive Relevance significantly outperforms both engagement signals and other survey measures in predicting next-day retention, especially for users with limited historical engagement. We develop a production-ready proxy model that integrates Retentive Relevance into the final stage of a multi-stage ranking system on a social media platform. Calibrated score adjustments based on this model yield substantial improvements in engagement, and retention, while reducing exposure to low-quality content, as demonstrated by large-scale A/B experiments. This work provides the first empirically validated framework linking content-level user perceptions to retention outcomes in production systems. We offer a scalable, user-centered solution that advances both platform growth and user experience. Our work has broad implications for responsible AI development.
\end{abstract}

%\ccsdesc[500]{Human-centered computing~Empirical studies in HCI}
%\ccsdesc[500]{Information systems~Social recommendation}
%\ccsdesc[500]{Information systems~Personalization}
%\ccsdesc[500]{Information systems~Content ranking}
%\ccsdesc[500]{Information systems~Recommender systems}

%% Keywords. The author(s) should pick words that accurately describe
%% the work being presented. Separate the keywords with commas.

\maketitle

\section{Introduction}
Recommendation systems are the backbone of modern digital platforms, guiding users through vast content libraries every day~\cite{ricci2022recommender}. Yet, the core challenge remains: how do we transform sparse, noisy engagement signals into reliable predictions of user preferences and long-term satisfaction~\cite{herlocker2004evaluating}? Most large recommendation systems rely on user engagement signals such as clicks, likes, comments and dwell time, operating under the assumption that these actions accurately reflect user interests and will result in future engagement and retention~\cite{hu2008collaborative,koren2009matrix}.
Yet, this engagement-centric approach is fundamentally limited. Large-scale studies reveal that users who interact with content do not always want more of the same~\cite{sharma2013social,hasan2024review}, and engagement signals are systematically biased toward popular items, often missing users' latent interests~\cite{abdollahpouri2017controlling,raza2024comprehensive}. This disconnect is especially problematic when optimizing for long-term effectiveness and retention, as short-term engagement frequently fails to predict sustained platform usage~\cite{jannach2015recommenders,gomez2015netflix,xue2025auro}.

To address these limitations, survey-based feedback has emerged as a promising alternative, offering direct insights into user preferences~\cite{mcnee2006accurate,pu2011user,hasan2024review,lv2025}. However, existing survey measures focus on capturing the immediate value of the recommendation for the user, lacking the forward-looking perspective required to optimize for retention. These retrospective measures, while informative, do not capture the behavioral intentions that drive users to return to platforms and underpin long-term effectiveness.

To bridge this gap, we introduce \textbf{Retentive Relevance}, a novel content-level survey measure designed to capture users' \textit{intent to return} for similar content. By asking users immediately after a recommendation, ``How likely or unlikely are you to return to [platform] to view more posts like this?'', Retentive Relevance directly measures the value of a recommendation as it relates to users' intent to return, while maintaining the clarity and interpretability of survey-based self-reported feedback.

Our comprehensive evaluation—encompassing offline analyses, large-scale production deployments, and A/B experiments— demonstrates that \textit{Retentive Relevance} consistently surpasses both traditional engagement signals and alternative survey measures in predicting next-day retention. This leads to improved user retention, increased engagement, and measurable gains in content quality and integrity metrics. Notably, our approach is particularly effective for low-signal users, where conventional metrics are sparse or unreliable.
The key contributions of this paper are:

\begin{itemize}
    \item \textbf{Novel survey construct with predictive validity for retention:} Retentive Relevance is the first content-level survey measure empirically validated to predict next-day user retention in recommendation systems.
    \item \textbf{Complete operational framework:} We present an end-to-end methodology and framework— from survey design to production model deployment— for designing, validating, and operationalizing Retentive Relevance at scale in recommendation systems.
    \item \textbf{Large-scale experimental validation:} We provide robust evidence from live A/B experiments deployed in a large social media platform demonstrating that Retentive Relevance drives significant improvements in user retention, engagement, and content quality.
    \item \textbf{Theoretical and practical insights:} We highlight the superior predictive power of forward-looking behavioral intent, and strong results in quality improvements offering implications for the design of responsible AI systems.
\end{itemize}

The remainder of this paper is organized as follows. Section~2 reviews related work and situates our contribution within the literature. Section~3 details our survey design, data collection, and bias correction methodology. Section~4 summarized the findings on the relationships between Retentive Relevance and other survey measures and engagement signals. Section~5 presents our offline retention modeling results and compares predictive performance of Retentive Relevance with other alternative survey measures. Section~6 covers our approach to building a proxy based on survey, integrating into ranking production system and results of our online A/B testing. Section~7 discusses implications and future directions.

\section{Related Work}
We structure our review of related work around four key areas: user feedback in recommendation systems, survey-based feedback mechanisms, retention prediction and long-term value, and methods for cold-start and low-signal users. This structure clarifies the landscape and highlights how our work advances the field.

\textbf{User Feedback in Recommendation Systems.}
Recommendation systems utilize both implicit feedback (e.g., clicks, dwell time~\cite{rendle2009bpr}) and explicit feedback (e.g., ratings, thumbs up/down~\cite{resnick1994grouplens,adomavicius2012improving}). While implicit signals are abundant, they are often noisy and biased~\cite{marlin2007collaborative,joachims2005accurately}. Explicit feedback provides more direct signals but is less frequent and can be affected by response and popularity bias~\cite{konstan2012recommender}. Most prior work focuses on immediate reactions to content, limiting the ability to predict long-term engagement~\cite{chen2019serendipity,zhang2014explicit}.

\textbf{Survey-Based Feedback Mechanisms.}
Within explicit feedback, survey-based methods have gained prominence for their ability to directly capture user satisfaction and interest~\cite{knijnenburg2012explaining,raza2024comprehensive,hasan2024review}. Surveys can complement behavioral metrics, address data sparsity, and improve model explainability~\cite{lv2025,covington2016deep,butmeh2024hybrid}. Research in this area explores optimal survey design, question framing, and timing~\cite{kumar2005model,liu2010personalized}. However, most surveys are retrospective, evaluating the current consumption value of content rather than capturing forward-looking intent, a gap our work aims to address.

\textbf{Retention Prediction and Long-Term Effectiveness.}
As digital platforms increasingly prioritize sustainable growth, accurately predicting user retention has emerged as a central challenge~\cite{jhawar2023quantifying, sun2022surrogate}. Recent research has introduced a range of advanced techniques to address this problem. Reinforcement learning frameworks have been developed to optimize cumulative long-term rewards~\cite{zheng2018drn}, while causal inference methods help disentangle the effects of specific content on user retention~\cite{schnabel2016recommendations}. Graph neural networks further enable the modeling of complex user-item-time interactions~\cite{wu2019session}. In addition, multi-task and sequential modeling approaches have been proposed to balance short- and long-term objectives~\cite{wang2018dkn,li2017neural}.
Adaptive retention optimization frameworks~\cite{xue2025auro} and generative flow networks~\cite{liu2024retention,liu2024sequential} have demonstrated significant improvements in next-day return prediction and overall engagement, with large-scale deployments validating the practical impact of retention-focused systems~\cite{cai2023reinforcing}. However, despite these advances, most existing methods continue to rely on noisy engagement signals and often lack explainable, recommendation-level approaches that can bridge the temporal gap between immediate user actions and future behavior.

\begin{table*} [!htbp]
    \centering
    \begin{tabularx}{\textwidth}{p{3.5cm}Xp{2cm}}
        \hline
        \textbf{Name (Construct)} & \textbf{Survey Question} & N (Sample size)\\
        \hline
        \textbf{Retentive Relevance} \newline (Likelihood to return) &
        \makecell[l]{How likely or unlikely are you to return to [Platform] to view more posts like this? \\
        Very likely, Likely, Neither likely or unlikely, Unlikely, Very unlikely}  & $N = 63{,}708$ \\
        \hline
        \textbf{Interest Matching} \newline (Interest relevance) &
        \makecell[l]{To what extent does this video match your interests? \\ A great deal, A lot, A moderate amount, A little, Not at all} & $N = 58{,}872$ \\
        \hline
        \textbf{Worth Your Time} \newline (Recommendation value) &
        \makecell[l]{Was this video worth your time? \\
        Completely, Mostly, Somewhat, Barely, Not at all} & $N = 76{,}263$ \\
        \hline
    \end{tabularx}
    \caption{"Retentive Relevance" was compared with two other survey measures. Each survey was administrated under equal conditions but separately. Data collection occurred between December 2024 and January 2025 across 18 countries on a large social media platform targetted to personalized video recommendation feed.}
    \label{tab:survey_questions}
\end{table*}

\textbf{Cold-Start and Low-Signal Users.}
The cold-start problem remains a fundamental challenge in personalization, particularly for new users or those with limited interaction history~\cite{schein2002methods}. Collaborative filtering methods are especially vulnerable to data sparsity, while content-based approaches may fail to capture valuable collaborative signals~\cite{burke2002hybrid}. Hybrid models attempt to mitigate these issues by integrating multiple signal types, but they often still rely heavily on noisy implicit feedback~\cite{adomavicius2005toward}.
Recent advances have explored meta-learning, few-shot, and transfer learning techniques to address cold-start and low-signal scenarios~\cite{vartak2017meta,lee2019melu,elkahky2015multi}. Additionally, large language models and graph-based methods have shown promise in extracting richer representations from auxiliary data~\cite{bei2025coldstart,li2024neural}. However, most focus on auxiliary data, more prone to accuracy issues, rather than capturing ground truth preferences via direct user feedback.
Survey-based methods offer a distinct advantage in cold-start contexts by enabling the immediate collection of explicit user preferences, even in the absence of substantial behavioral history.

\textbf{Our Contribution.}
This paper advances the field at the intersection of survey-based ground truth signal collection, retention prediction, and production-scale integration with recommendation systems. We introduce an end-to-end framework that is rigorously validated through large-scale offline analyses and live online experiments. Our approach enables the direct integration of long-term user intent into algorithmic optimization, providing a scalable and interpretable signal for improving user retention. Beyond technical impact, our framework offers practical implications for broader responsible AI systems, supporting more user-aligned algorithmic systems and sustainable platform growth.

\section{Survey Implementation, Data Collection and Bias Correction}
In this section we discuss the theoretical foundation for this work and the approach for developing the survey instrument, validating it, collecting data and correcting bias.

\textbf{Theoretical Foundation-}
We designed Retentive Relevance to capture users' forward-looking intentions to return to a recommendation platform based on the value they perceive in the content. Unlike other survey-based measures that focus on immediate value or interest relevance (See Table~\ref{tab:survey_questions}), Retentive Relevance specifically targets the antecedents of retention behavior. This approach is grounded in the Theory of Planned Behavior~\cite{ajzen1991theory}, which posits that behavioral intentions are the strongest predictors of actual behavior, as well as established research on behavioral intention measurement~\cite{fishbein1975belief}.
The key theoretical distinction between Retentive Relevance and other constructs lies in its temporal orientation and behavioral specificity. For example, Interest Matching (see Table~\ref{tab:survey_questions}) captures cognitive alignment between content and user preferences, while Worth-Your-Time assesses retrospective value. In contrast, Retentive Relevance explicitly probes the likelihood of future behavior, aligning more closely with the retention outcomes we aim to predict.

\textbf{Survey Instrument Development-}
Following best practices in survey development~\cite{tourangeau2000psychology, willis2005cognitive, groves2009survey}, we employed a theory-driven, backwards-design approach. The survey item was formulated as: \textit{"How likely or unlikely are you to return to [platform] to view more posts like this?"} where [platform] refers to the large-scale social media app where the survey was conducted. Responses were collected on a balanced 5-point Likert scale ranging from Very unlikely (1) to Very likely (5), with a neutral midpoint.
The question wording was carefully crafted to specify the behavioral target ("return to [platform]"), clarify content specificity ("posts like this"), and capture likelihood rather than certainty, acknowledging the inherent uncertainty in predicting future behavior.

\textbf{Construct Validation Protocol-}
To establish content validity~\cite{brown2015confirmatory} and ensure comprehension across diverse user populations, we conducted cognitive testing following standardized protocols~\cite{willis2005cognitive, tourangeau2000psychology}. Literature suggests that 5--12 participants are sufficient to identify most comprehension issues~\cite{willis2005cognitive}. We recruited $N = 8$ participants from the United States, stratified by gender (50\% female), age (18--24: 25\%, 25--40: 50\%, 41--65: 25\%), and platform usage (active vs. infrequent users: 50\%/50\%). Each think-aloud session lasted approximately 30 minutes.
Using standardized cognitive interviewing methods~\cite{tourangeau2000psychology}, we systematically assessed four cognitive processes underlying survey response. We evaluated 1) \textit{Comprehension} by asking participants ``What does this question mean to you?'' to assess understanding of the forward-looking, behavioral nature of the question. 2) \textit{Retrieval processes} were examined through ``What specific content were you thinking about?'' to evaluate whether participants referenced the intended recommendation. 3) \textit{Judgment formation} was assessed by asking ``How did you decide on your rating?'' to examine the decision-making process and influencing factors. Finally, 4) \textit{Response mapping} was evaluated through ``Was it easy to select from the provided options?'' to assess the appropriateness of the scale and response burden.
Results indicated consistent understanding of the Retentive Relevance construct, with an average inter-rater agreement of 87.5\% on key comprehension items. Participants reliably distinguished Retentive Relevance from alternative measures (e.g., Interest Matching and Worth Your Time) with 87.5\% accuracy, as measured by the proportion who consistently identified the intended construct in comparison scenarios. Importantly, participants demonstrated clear conceptual differentiation between immediate content evaluation (``Was this good?'') and future behavioral intention (``Will I come back for more like this?''), supporting the theoretical basis of our construct.

\begin{figure}[!htb]
\centering
\includegraphics[width=0.5\linewidth]{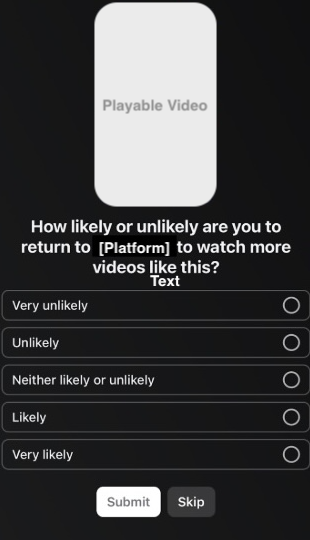}
\caption{Schematic representation of the Retentive Relevance survey implementation. The interface maintains visual reference to the content being evaluated while capturing forward-looking behavioral intentions. "Platform" was replaced with the name of the social media app where the survey was deployed.}
\label{fig:schematic}
\end{figure}

\textbf{Survey Implementation-}
Surveys were implemented as a contextual overlay, appearing immediately after a video recommendation to minimize recall bias and maximize ecological validity. This timing ensures that users evaluate content while their experience and emotional response are still salient, reducing the cognitive burden and potential bias of retrospective evaluation~\cite{tourangeau2000psychology}.
The survey interface displayed a playable video thumbnail above the question (see example in Figure~\ref{fig:schematic}), allowing users to reference the content while responding. To mitigate response bias~\cite{groves2009survey}, we incorporated several design features including randomized response order to counteract order effects, balanced scale anchors to prevent directional bias, and a neutral midpoint to accommodate genuine ambivalence.
Survey triggers were programmed to appear randomly across all video recommendations by feed position and regardless of user interaction (e.g. watched, engaged or skipped), ensuring unbiased sampling across the content valuation spectrum and preventing systematic exclusion of skipped content.
Survey questions and response options were translated into users' local languages following established internationalization practices, with back-translation validation to ensure construct equivalence. The implementation was designed to ensure that data collection did not significantly disrupt the experience and always provided the option to skip the survey.

\textbf{Data Collection-}
We collected survey responses $N = 63{,}708$ for Retentive Relevance, $N = 58{,}872$ for
Interest Matching, and $N = 76{,}263$ for Worth Your Time under equivalent conditions and statistical treatment between December 2024 and January 2025 across 18 countries using stratified sampling by user engagement levels (active vs. less active users). For
each survey response, we collected multi-level features at the user level (e.g. historical engagement, same-day engagement, next day engagement and demographics), content level (e.g. content topic, content age, overall content level engagements and popularity) and user-content level interactions (e.g. likes, comments, shares, watch time, skip, etc.).

\textbf{Bias Correction-}
To address nonresponse bias, we implemented covariate balancing propensity scores following established practices~\cite{schnabel2016recommendations,joachims2017unbiased}. Our propensity score model incorporated user demographics (age cohorts, geographic regions, platform tenure), behavioral patterns (engagement and consumption levels), and platform features (tenure on platform). The Covariate Balancing Propensity Score (CBPS) optimization balances covariate distributions while estimating propensity scores:

\begin{equation}
\mathbb{E}[\pi(X_i)(1-\pi(X_i))X_i] = \frac{1}{n}\sum_{i=1}^n (Z_i - \pi(X_i))X_i = 0
\end{equation}

where $\pi(X_i)$ represents the propensity to respond to surveys and $Z_i$ indicates survey completion.
Post-weighting evaluation achieved standardized mean differences $|SMD| < 0.1$ across all covariates~\cite{austin2009balance}, with trimming applied for extreme propensity scores following established practices~\cite{crump2009dealing}.

\section{Retentive Relevance vs. Alternative Surveys and Engagement Signals}
To ensure that Retentive Relevance both captures the value of recommendations and remains distinct from other survey and engagement measures, we rigorously validated its psychometric properties —specifically, its convergent and discriminant validity— using established principles.

\textbf{Convergent Validity and Relationships with Other Survey Measures.}
Convergent validity assesses whether measures that are theoretically related exhibit strong positive correlations, while still maintaining distinct conceptual boundaries~\cite{campbell1959convergent,nunnally1994psychometric}. Following established validation protocols~\cite{cohen1988statistical,nunnally1994psychometric}, we evaluated convergent validity by analyzing correlations between user-level survey responses within similar content types.
To enable meaningful cross-sample comparisons, we computed the mean response for each measure within specific content categories at the user level. The resulting cross-sample correlations revealed significant positive associations among all measures, providing robust evidence for convergent validity. Notably, Retentive Relevance showed substantial correlations with Worth Your Time (r = 0.63, p < 0.001, 95\% CI [0.71, 0.75]) and Interest Matching (r = 0.58, p < 0.001, 95\% CI [0.66, 0.70]). These values fall within the optimal range for convergent validity~\cite{cohen1988statistical}, indicating meaningful conceptual overlap while remaining sufficiently below the threshold (r < 0.85) that would suggest redundancy~\cite{kline2015principles}.

\textbf{Discriminant Validity: What Makes Retentive Relevance Distinct.}
Discriminant validity requires that measures of theoretically distinct constructs display different response patterns across varied contexts~\cite{campbell1959convergent}. We assessed this by examining how our survey measures differentiated between types of recommendation value across content categories.
Our analysis revealed clear contextual differences in the relationships between measures. For content with immediate utilitarian or emotional value (e.g., motivation, learning, DIY), Retentive Relevance correlated more strongly with Worth Your Time (mean r = 0.69) than with Interest Matching (mean r = 0.55). In contrast, for interest-driven content (e.g., celebrities, technology, fashion), Retentive Relevance was more closely aligned with Interest Matching (mean r = 0.65) than with Worth Your Time (mean r = 0.51). This pattern suggests that Retentive Relevance adapts to different content contexts, capturing a broader spectrum of recommendation value.
We further validated these differences using Fisher's z-transformation to compare correlation coefficients across content types. All observed differences were statistically significant (z > 2.58, p < 0.01), confirming that the measures respond systematically differently to distinct content topics.

\begin{figure}[!t]
\centering
\includegraphics[width=1\linewidth]{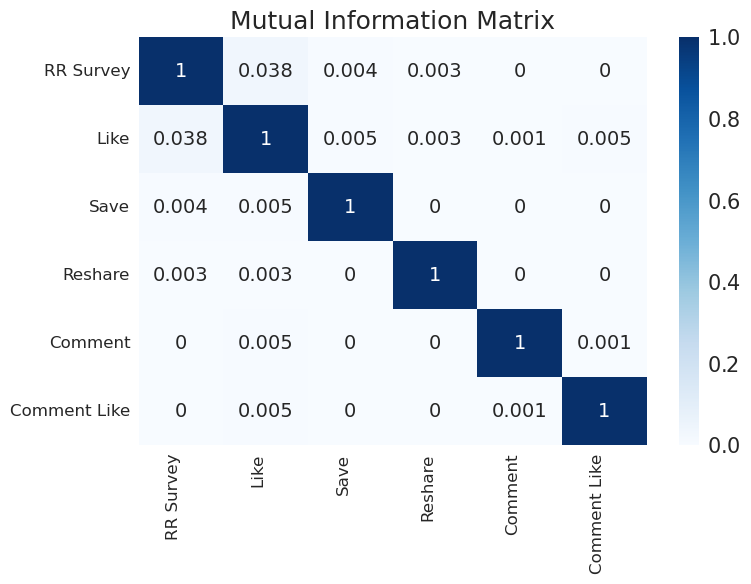}
\caption{Heatmap of Mutual Information Matrix shows that Retentive Relevance captures information about recommendation that is distinct from engagement signals.}
\label{fig:mutual_info}
\end{figure}

\textbf{Orthogonality to Existing Engagement Signals.}
To establish Retentive Relevance as a valuable and actionable signal for recommendation systems, it is crucial to demonstrate that it provides incremental information beyond what is captured by traditional engagement signals. We quantified the dependence between Retentive Relevance and standard engagement signals using mutual information, which measures how much knowing one variable reduces uncertainty about the other.
As shown in Figure~\ref{fig:mutual_info}, the heatmap of mutual information coefficients reveals that Retentive Relevance consistently exhibits low mutual information with all traditional engagement signals (MI < 0.15 for all signals), indicating substantial independence from behavioral indicators. This finding underscores that user-stated intentions, as measured by Retentive Relevance, provide distinct and complementary information that cannot be inferred from observed engagement alone.

\section{Predictive Performance of Survey-Based Signals in Retention Models}
\begin{table*}[!htb]
\centering
\caption{Predictive performance for next-day retention models shows that Retentive Relevance yields substantial and statistically significant gains in both accuracy and ROC AUC, while models with alternative survey measures do not show any statistically significant improvements.}
\label{tab:predictive_performance}
\begin{tabular}{l@{\hspace{3em}}cc@{\hspace{4em}}cc}
\toprule
& \multicolumn{2}{c}{\textbf{Overall Sample}} & \multicolumn{2}{c}{\textbf{Low-Signal Users}} \\
\cmidrule(lr){2-3} \cmidrule(lr){4-5}
\textbf{Model} & \textbf{Accuracy (\%)} & \textbf{ROC AUC} & \textbf{Accuracy (\%)} & \textbf{ROC AUC} \\
\midrule
Baseline (No Survey) & 78.0 $\pm$ 0.3 & 0.830 $\pm$ 0.005 & 73.0 $\pm$ 1.3 & 0.630 $\pm$ 0.013 \\
+ Retentive Relevance & \textbf{83.0 $\pm$ 0.3***} & \textbf{0.860 $\pm$ 0.005***} & \textbf{76.0 $\pm$ 1.5***} & \textbf{0.700 $\pm$ 0.025***} \\
+ Worth Your Time & 78.0 $\pm$ 0.4 & 0.828 $\pm$ 0.006 & 73.2 $\pm$ 1.4 & 0.632 $\pm$ 0.015 \\
+ Interest Matching & 78.2 $\pm$ 0.3 & 0.838 $\pm$ 0.005 & 73.1 $\pm$ 1.3 & 0.635 $\pm$ 0.014 \\
\bottomrule
\end{tabular}
\begin{tablenotes}
\centering
\small
\item Results show mean $\pm$ 95\% CI from stratified 10-fold cross-validation. ***$p < 0.001$ compared to baseline via paired t-test.
\item Bold indicates best performance for each metric.
\end{tablenotes}
\end{table*}

Having established that Retentive Relevance is a valid measure of personalized recommendation quality-demonstrating convergent validity while remaining distinct from other survey and engagement measures—we now evaluate its predictive power for next-day retention behavior. This analysis is designed to establish the behavioral validity of Retentive Relevance by comparing its predictive performance against alternative survey measures.

\textbf{Modeling Approach.}
We formulate next-day retention prediction as a binary classification problem to assess the incremental value of survey responses. For each user $i$, the retention outcome $y_i \in \{0, 1\}$ indicates whether the user returns the following day, where $y_i = 1$ represents retention defined as video recommendation views exceeding the 5th percentile threshold of active user distributions. This operationalization distinguishes genuine retention behavior from accidental or minimal platform engagement.

We construct feature vectors $\mathbf{x}_i \in \mathbb{R}^d$ that capture multiple dimensions of user behavior and context. The feature vector is composed of five distinct components: $\mathbf{x}_i = [\mathbf{h}_i, \mathbf{r}_i, \mathbf{u}_i, \mathbf{c}_i, \mathbf{d}_i]$, where $\mathbf{h}_i$ represents historical engagement features aggregated over 28 days, $\mathbf{r}_i$ captures real-time signals including same-day activity patterns, $\mathbf{u}_i$ encompasses user-content interactions through both explicit and implicit feedback, $\mathbf{c}_i$ includes content metadata such as topic classification and creator characteristics, and $\mathbf{d}_i$ provides demographic and usage controls including age cohort and platform tenure.

We employ XGBoost gradient boosting classifiers optimized for log-loss, leveraging their robust performance with heterogeneous features and built-in regularization capabilities. The model prediction is formulated as:

\begin{equation}
\hat{y}_i = \sigma\left(\sum_{k=1}^{K} f_k(\mathbf{x}_i)\right)
\end{equation}

where $f_k$ represents the $k$-th tree in the ensemble, $K$ denotes the total number of trees, and $\sigma(\cdot)$ is the sigmoid function mapping ensemble outputs to probability space.
To assess the incremental value of each survey measure $s \in \{\text{RR}, \text{WYT}, \text{IM}\}$ (Retentive Relevance, Worth Your Time, Interest Matching), we construct paired model comparisons:

\begin{align}
M_{\text{baseline}}: &\quad P(y_i = 1 | \mathbf{x}_i) \\
M_{\text{augmented}}: &\quad P(y_i = 1 | \mathbf{x}_i, s_i)
\end{align}

where $s_i$ represents the survey response for user $i$. This paired design enables direct quantification of survey signal contributions while controlling for all other predictive factors.

We employ 10-fold cross-validation to maintain outcome class proportions across folds. For each fold $j \in \{1, \ldots, 10\}$, we compute performance metrics $\mathcal{M}_j$ including accuracy and ROC AUC for both baseline and augmented models. The incremental predictive value is quantified as the mean performance difference across folds:

\begin{equation}
\Delta \mathcal{M} = \frac{1}{10} \sum_{j=1}^{10} \left(\mathcal{M}_j^{\text{augmented}} - \mathcal{M}_j^{\text{baseline}}\right)
\end{equation}

Statistical significance is assessed using paired t-tests across folds, with effect sizes calculated using Cohen's d. Bootstrap confidence intervals with 1000 iterations provide robust uncertainty estimates for performance improvements.

\textbf{Predictive Performance Results.}
Table~\ref{tab:predictive_performance} presents the cross-validated performance results for next-day retention prediction with and without the survey measures. The results demonstrate that Retentive Relevance provides substantial and statistically significant improvements in both accuracy and ROC AUC. For the overall sample, incorporating Retentive Relevance into the prediction model increased accuracy by 5.0 percentage points (from 78.0\% to 83.0\%) and ROC AUC by 0.030 points (from 0.830 to 0.860), with significant effect sizes (Cohen's d = 2.1, $p < 0.001$).

The predictive gains were more pronounced for low-signal users, i.e. those with limited historical engagement data. For this user segment, Retentive Relevance increased accuracy by 3.0 percentage points and ROC AUC by 0.070 points (Cohen's d = 3.2, $p < 0.001$). The magnitude of these gains is particularly meaningful in large-scale recommendation systems, where even modest percentage increases can translate to additional retained users, especially considering these effects result from a single recommendation interaction.
In contrast, neither Worth Your Time nor Interest Matching surveys provided significant predictive value for next-day retention, underscoring that Retentive Relevance captures unique behavioral intentions specifically relevant to retention decisions, rather than general content satisfaction or interest alignment captured by existing survey measures.

\begin{figure*}[!t]
\centering
\includegraphics[width=0.8\linewidth]{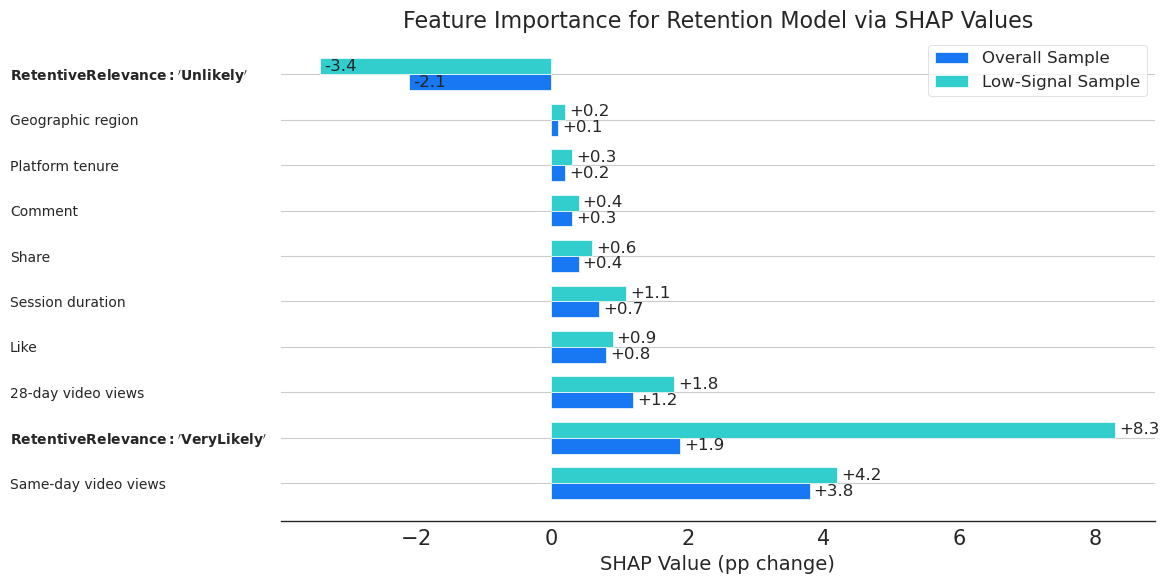}
\caption{Feature importance analysis via SHAP values shows that Retentive Relevance significantly improves retention prediction—especially for low-signal users and with survey signals proving more predictive than engagement signals.}
\label{fig:shap_analysis}
\end{figure*}

\textbf{Feature Importance and Model Interpretation.}
We conducted feature importance analysis using SHAP (SHapley Additive exPlanations) values~\cite{lundberg2017unified}, quantifying each feature's marginal contribution to individual predictions, expressed as percentage point changes in predicted retention probability(See Figure~\ref{fig:shap_analysis}).

For the general population, "Unlikely" Retentive Relevance responses emerge as the second most important negative predictor (-2.1 pp), ranking immediately after same-day engagement controls. This substantial negative impact validates the behavioral connection between stated intent and actual retention outcomes, demonstrating that users who express low likelihood of returning indeed exhibit lower likelihood of returning for video views the next day.
The effect becomes dramatically amplified for low-signal users, where "Very Likely" Retentive Relevance responses constitute the strongest positive predictor after controlling for same-day activity (+8.3 pp). This effect size substantially exceeds any traditional engagement factor, including likes, shares and comments. The magnitude of this impact underscores the particular value of direct intent measurement for users where behavioral signals are limited or unreliable.
Across both user populations, Retentive Relevance responses consistently demonstrate superior predictive importance compared to traditional engagement measures. This disparity indicates that direct user intent signals provide substantially more predictive information than preferences inferred from behavioral observation alone.

These results establish that Retentive Relevance provides both statistically significant and practically meaningful improvements in retention prediction, with effect sizes that justify the implementation costs of survey-based feedback collection in production recommendation systems. The superior performance compared to established survey measures demonstrates that Retentive Relevance captures unique aspects of user experience specifically relevant to retention behavior, establishing its criterion validity as a forward-looking measure of user intent.

\section{Production Integration and Online Evaluation}
Having established the predictive validity of Retentive Relevance through comprehensive offline analysis, we now describe the end-to-end process of operationalizing survey signals within large-scale production recommendation systems. Our framework consists of three key phases: (1) development of production-ready proxy models that translate survey insights into real-time predictions, (2) integration of these predictions into existing ranking infrastructure through calibrated score adjustments, and (3) validation through large-scale online experimentation.

\textbf{Survey Signal Proxy Model.}
We formulate survey signal prediction as a binary classification problem to estimate user retention intent for unseen user-item pairs. Let $\mathcal{U}$ and $\mathcal{V}$ denote the sets of users and items, respectively. For any user-item pair $(u,v) \in \mathcal{U} \times \mathcal{V}$, we aim to predict the probability that user $u$ would express positive retention intent for item $v$.
Given the 5-point Likert scale survey responses, we adopt a binary classification framework where positive intent corresponds to "Likely" or "Very Likely" responses (ratings 4-5), and negative intent corresponds to "Unlikely" or "Very Unlikely" responses (ratings 1-2). Neutral responses (rating 3) are excluded from training, as their inclusion decreased discriminative performance by 2.3\% AUC.
The proxy model is formulated as a logistic regression classifier optimized for production deployment:

\begin{equation}
P(\text{RR}_{u,v} = 1 | \mathbf{x}_{u,v}) = \sigma(\mathbf{w}^T \mathbf{x}_{u,v} + b)
\end{equation}

where $\sigma(\cdot)$ is the sigmoid function, $\mathbf{w}$ represents learned weights, $b$ is the bias term, and $\mathbf{x}_{u,v} \in \mathbb{R}^d$ is the feature vector for user-item pair $(u,v)$.
The feature vector incorporates multiple signal categories following established practices~\cite{covington2016deep,cheng2016wide}:

\begin{equation}
\mathbf{x}_{u,v} = [\mathbf{p}_{u,v}, \mathbf{e}_{u}, \mathbf{c}_v, \mathbf{i}_{u,v}, \mathbf{n}_{u,v}]
\end{equation}

where $\mathbf{p}_{u,v}$ represents behavioral prediction scores including learned probabilities for engagement actions, $\mathbf{e}_u$ captures temporal engagement rate features, $\mathbf{c}_v$ includes content metadata, $\mathbf{i}_{u,v}$ represents user-content interaction patterns, and $\mathbf{n}_{u,v}$ encompasses negative feedback indicators.
The model is trained to minimize regularized logistic loss:

\begin{equation}
\mathcal{L}(\mathbf{w}, b) = -\frac{1}{N} \sum_{i=1}^{N} \left[ y_i \log p_i + (1-y_i) \log(1-p_i) \right] + \lambda \|\mathbf{w}\|_2^2
\end{equation}

where $N$ is the number of training samples, $y_i \in \{0,1\}$ is the binary survey label, $p_i = P(\text{RR}_{u_i,v_i} = 1 | \mathbf{x}_{u_i,v_i})$, and $\lambda$ is the L2 regularization parameter.

\textbf{Ranking Integration Architecture.}
Survey signal predictions are integrated into the final ranking stage of our multi-stage recommendation system on a large social media platform serving video recommendations. Let $\text{score}_{\text{base}}(u,v)$ denote the baseline ranking score for user $u$ and item $v$. The survey-augmented ranking score is computed as:

\begin{equation}
\text{score}_{\text{final}}(u,v) = \text{score}_{\text{base}}(u,v) + \text{boost}(u,v) + \text{demote}(u,v)
\end{equation}

where boost and demotion factors are defined as:

\begin{align}
\text{boost}(u,v) &= \alpha \cdot \mathbb{I}[\hat{p}_{u,v} > \tau_{\text{boost}}] \\
\text{demote}(u,v) &= -\beta \cdot \mathbb{I}[\hat{p}_{u,v} < \tau_{\text{demote}}] \cdot (\tau_{\text{demote}} - \hat{p}_{u,v})
\end{align}

Here, $\hat{p}_{u,v}$ is the predicted retention intent probability, $\alpha > 0$ and $\beta \in (0,1)$ are tunable parameters, and $\tau_{\text{boost}}$ and $\tau_{\text{demote}}$ are precision-calibrated thresholds with $\tau_{\text{demote}} < \tau_{\text{boost}}$.

Threshold calibration follows a data-driven approach optimizing for precision and coverage. The boost threshold $\tau_{\text{boost}}$ achieves 80\% positive precision at $\hat{p}_{u,v} > 0.76$, ensuring only high-confidence positive predictions receive ranking boosts. The demotion threshold $\tau_{\text{demote}}$ targets 60\% negative precision at $\hat{p}_{u,v} < 0.38$, balancing sensitivity and specificity. Items with predicted probabilities in the neutral zone $[\tau_{\text{demote}}, \tau_{\text{boost}}]$ receive no treatment, maintaining ranking stability for uncertain predictions.

\textbf{Online Experimental Results.}
We conducted large-scale online A/B experiments on a major social media platform with personalized video recommendations. The experimental design follows established best practices for recommendation system evaluation, incorporating comprehensive statistical rigor and multiple validation approaches.

We evaluated system performance across three primary metric categories: (1) User retention measured by sessions per user, (2) Engagement activity measured by metrics such as communication activity, like rates, and skip rates; (3) Content quality and integrity, measured through prevalence of reported content, negative feedback indicators, and established metrics based on quality and integrity classifiers.
All metrics were tracked continuously throughout the experiment window, with statistical significance assessed using two-sample t-tests and effect sizes calculated using Cohen's d. Bootstrap confidence intervals provided robust uncertainty estimates for observed differences.

Table~\ref{tab:metric_changes} summarizes the statistically significant changes observed across key platform metrics during the 14-day experimental period. The results demonstrate consistent improvements across user engagement, retention, and content quality metrics.

\begin{table}[!htb]
\centering
\caption{Results from online A/B testing show that integrating Retentive Relevance into ranking leads to significant improvements in retention, engagement, and content quality metrics.}
\label{tab:metric_changes}
\begin{threeparttable}
\begin{tabular}{p{2cm}p{2.8cm}c}
\toprule
\textbf{Category} & \textbf{Metric} & \textbf{Change (\% $\Delta$ ± 95\% CI)} \\
\midrule
\multirow{1}{*}{Retention} & Sessions per User & +0.030 ± 0.026 \\
\midrule
\multirow{3}{*}{Engagement} & Communication Activity & +0.052 ± 0.039 \\
& Like Rate & +0.169 ± 0.100 \\
& Skip Rate & --0.188 ± 0.085 \\
\midrule
\multirow{5}{*}{Content Quality} & Reported Content & --1.36 ± 0.11 \\
& Negative Feedback & --1.527 ± 0.095 \\
& "Not Interested" Feedback & --2.6 ± 1.3 \\
& Reports to Likes Ratio & --0.825 ± 0.075 \\
\bottomrule
\end{tabular}
\begin{tablenotes}
\small
\item All changes reported as percentage point differences with 95\% confidence intervals. Negative values indicate reductions; positive values indicate increases. All reported changes are statistically significant at $p < 0.05$.
\end{tablenotes}
\end{threeparttable}
\end{table}

The treatment group showed enhanced user interaction patterns, with communication activity increasing by 0.052 percentage points (±0.039), like rates by 0.169 percentage points (±0.100), and skip rates decreasing by 0.188 percentage points (±0.085). Note that users often show their lack of interest in content via skip. Most critically, sessions per user increased by 0.030 percentage points (±0.026), reflecting improved retention.

Additionally, Retentive Relevance integration yielded improvements in content quality metrics. Prevalence of reported content decreased by 1.36 percentage points (±0.11), negative feedback declined by 1.527 percentage points (±0.095), and "not interested" signals dropped by 2.6 percentage points (±1.3). These reductions demonstrate the system's enhanced ability to identify and demote low quality content while improving user experience.
The results demonstrate that optimizing for Retentive Relevance creates natural alignment between improved user experience, platform growth as well as improved content quality.

\section{Discussion and Implications}
Building on our results, we now explore the broader implications, limitations, and future directions of Retentive Relevance for recommendation systems and AI applications.

\textbf{User-Centered Paradigm: Shifting the Foundation.}
Our work establishes a user-centered paradigm for recommender systems by empirically validating the connection between content-level user perceptions and retention outcomes, aligning with the growing emphasis on intent-based AI systems.
We show that survey responses capturing users' future intent are stronger predictors of actual behavior than traditional survey or engagement signals, grounding this finding in the Theory of Planned Behavior~\cite{ajzen1991theory}, which posits that behavioral intentions are the strongest predictors of actual behavior. This theoretical foundation distinguishes our approach from content-based and collaborative filtering methods that rely heavily on past interactions~\cite{xu2025enhancing}. The unique, orthogonal predictive power of Retentive Relevance---distinct from existing engagement metrics---demonstrates that intent-based feedback reveals fundamentally different aspects of user preferences, addressing the limitation that users who interact with content do not always want more of the same~\cite{hasan2024review,sharma2013do}.
This insight empowers platforms to move beyond optimizing for short-term engagement, enabling a focus on long-term value and sustained user retention, identified in recent literature on sustainable AI deployment and user experience optimization.

\textbf{Practical Impact and Industry Applications.}
We present an end-to-end framework, from survey design to production deployment, validated through large-scale online A/B testing. This production-ready approach is broadly applicable to other AI systems beyond recommendations, wherever user feedback and intent can help optimize or calibrate complex models. We show that optimizing for user intent drives simultaneous improvements in platform retention, engagement, and content quality metrics. These results demonstrate that user-centered optimization can resolve longstanding trade-offs between growth and responsibility---a critical consideration as organizations scale AI across multiple departments and business processes~\cite{ibm2024scaling}. The measured improvements in content integrity metrics provide empirical evidence that intent-based optimization creates natural alignment between user experience and platform growth and quality.

\textbf{Implications for Responsible AI Systems.}
Incorporating user feedback directly into AI systems has significant implications for responsible AI development. In our approach users directly express their intent, making algorithmic decisions more interpretable compared to systems that infer preferences from opaque behavioral signals.
By enabling users to express their intent and preferences, recommendation algorithms become more transparent, accountable, and aligned with individual values. Ultimately, user-centered feedback mechanisms represent a step toward building AI systems that are not only effective but also align with users' long-term interests and values.

\textbf{Limitations and Future Directions.}
While Retentive Relevance demonstrates strong effectiveness, several limitations present opportunities for future research. Currently, our approach captures the value of a single recommendation interaction, missing the broader context of user sessions and sequence of recommendations. Future work could explore session-level and experience survey designs that can be used directly as optimization objectives rather than as additive signals, potentially incorporating advances in sequential modeling and multi-task learning~\cite{raza2024comprehensive}.
Longitudinal tracking can further illuminate the evolution of user intent over time, addressing how preferences shift across different contexts and temporal patterns.
Additionally, expanding the framework to cross-modal recommendations and other AI systems could broaden its
applicability.

\section{Conclusion}
In this paper, we introduce Retentive Relevance—a novel, survey-based measure that advances recommendation system evaluation from retrospective satisfaction to forward-looking user intent. By directly capturing users’ likelihood to return, we demonstrate that Retentive Relevance outperforms both traditional engagement signals and alternative survey measures in predicting user return to the platform. Integrating Retentive Relevance into ranking and validating it through online A/B testing, we show that it provides valuable additional signal on user preferences and drives improvements in retention, engagement, and content quality at scale. We propose that Retentive Relevance serves as a scalable, model-agnostic approach that bridges user perception research and production, setting a new standard for responsible, user-centered AI personalization.

\begin{acks}
We acknowledge the survey participants who provided valuable feedback that made this research possible.
\end{acks}

\bibliographystyle{ACM-Reference-Format}
\bibliography{references}

\end{document}